\documentclass[aps,prb,twocolumn,showpacs,superscriptaddress]{revtex4}
\usepackage{bm,amsmath,amssymb,latexsym,mathrsfs,enumerate}
\usepackage[mathcal]{euscript}
\usepackage{graphicx, subfigure}
\newcommand{\figwidth}{0.47\textwidth}

\begin{document}

\title{Non-linear dynamics and two-dimensional solitons
for spin $ S=1$ ferromagnets with biquadratic exchange}

\author{B. A. Ivanov}
\email{bivanov@i.com.ua} \affiliation{Institute of Magnetism, 03142
Kiev, Ukraine} \affiliation{National Taras Shevchenko University of
Kiev, 03127 Kiev, Ukraine}

\author{ A. Yu. Galkin}
\affiliation{Institute of Metal Physics, 03142 Kiev, Ukraine}
\affiliation{Institute of Magnetism, 03142 Kiev, Ukraine}

\author{R. S. Khymyn}
\affiliation{National Taras Shevchenko University of Kiev, 03127
Kiev, Ukraine} \affiliation{Institute of Magnetism, 03142 Kiev,
Ukraine}

\author{A. Yu. Merkulov}
\affiliation{FOM Institute for Atom and Molecular Physics,
Macromolecular Ion Physics Group, Kruislaan 407 1098 SJ Amsterdam,
The Netherlands.}

\date\today

\begin{abstract}
We develop a consistent semiclassical theory of spin dynamics for an
isotropic ferromagnet with a spin $ S=1$ taking into consideration
both bilinear and biquadratic over spin operators exchange
interaction. For such non-Heisenberg magnets,  a peculiar class of
spin oscillations and waves, for which the quantum spin expectation
value $ {\rm {\bf m}}=\langle{\rm {\bf S}}\rangle$ does not change
it direction, but changes in length, is presented. Such
``longitudinal'' excitations do not exist in regular magnets,
dynamics of which are described in terms of the Landau-Lifshitz
equation or by means of the spin Heisenberg Hamiltonian. We
demonstrate the presence of non-linear uniform oscillations and
waves, as well as self-localized dynamical excitations (solitons)
with finite energy. A possibility of excitation of such oscillations
by ultrafast laser pulse is discussed.
\end{abstract}



\maketitle
\section{Introduction} Magnetically ordered materials (magnets) are known
as essentially nonlinear systems.\cite{GurevMelk,Bar-springer}
Localized nonlinear excitations with finite energy, or solitons,
play an important role in description of nonlinear dynamics, in
particular, spin dynamics for low-dimensional magnets, with
different kind of magnetic order. To date, solitons in Heisenberg
ferromagnets, whose dynamics are described by the Landau--Lifshitz
equation for the constant-length magnetization vector, have been
studied in details, see for review
Refs.~\onlinecite{Bar-springer,Kosevich+All,MikStainer,BarIvKhalat,IvanovFNT05}.
In terms of microscopic spin models, this picture corresponds to the
exchange Heisenberg Hamiltonian, with the isotropic bilinear spin
interaction $ J{\rm {\bf S}}_1 {\rm {\bf S}}_2 $.\cite{SW} For a
spin of $ S>1/2$ the isotropic interaction is not limited by this
term and can include higher invariants such as $ ({\rm {\bf S}}_1
{\rm {\bf S}}_2 )^n$ with $ n$ up to $ 2S$. In particular, the
general isotropic model with the spin $ S=1$ and the nearest
neighbor interaction is described by the Hamiltonian
\begin{equation}
\label{eq1} \hat {\mathcal{H}}=-\sum\limits_{<i,j>} {\left[ {J\left(
{{\rm {\bf S}}_i {\rm {\bf S}}_j } \right) + K\left( {{\rm {\bf
S}}_i {\rm {\bf S}}_j } \right)^2} \right]}.
\end{equation}
Here, the constants $ J$ and $ K$ determine the spin-bilinear
(Heisenberg) and spin-biquadratic exchange interactions between
nearest neighbors $ \left\langle {i,j} \right\rangle $. This model
(\ref{eq1}) has been actively studied for the last two decades both
in view of description of usual crystalline magnets, see pioneering
articles Refs.~\onlinecite{ChenLevy,Matveev} and review articles
Refs.~\onlinecite{Nagaev88,LoktOstr94} and in application to
low-dimensional magnets, see recent  Refs.
\onlinecite{IvanovKol03,FathSolyom05}.

For the model (\ref{eq1}) the character of the ground state is more
complicated, than for Heisenberg magnets. It is determined by the
values of the parameters of bilinear and quadratic exchange, $ J$
and $ K$. In addition to the ferromagnetic phase, which is stable at
$ J>K$, $ J>0$, and the antiferromagnetic phase, which is stable
within the mean field approximation at $ J<K$, $ J<0$, two so-called
nematic phases (collinear and orthogonal, see,\cite{triplet}), are
realized for this model. For these nematic states, the quantum spin
expectation value $ \mathbf{m}=\langle {\rm {\bf S}}\rangle$ equals
zero, even at zero temperature. (Below for short we will refer to
the vector $ \mathbf{m}$ as magnetization). The areas of existing of
the nematic phases separate from both sides the domains of stability
for the ferromagnetic and antiferromagnetic phases. Interest to the
model (\ref{eq1}) increases in view of investigation of
multicomponent Bose-Einstein condensates of neutral atoms with
nonzero spins.\cite{Zhou}. At two chosen values of $ J/K$, namely,
at $ J=K$ and $ J=0$, the model (\ref{eq1}) has the symmetry
$\mbox{SU(3)}$, which is higher than the rotational symmetry
inherent to $ \mbox{SO}(3) \sim \mbox{SU}(2)$, and in a
one-dimensional case is exactly integrable. The latter is
interesting from the theoretical point of view.

A possibility to change the magnetization  $ \mathbf{m}$ in length
is an important peculiarity of the ferromagnetic phase in the model
(\ref{eq1}). It is worth noting, for the regular Landau-Lifshitz
equation, frequently employed for description of spin dynamics, the
magnetization length keeps constant. This property is associated
with the fact that the Landau-Lifshitz equations naturally emerge
within the approach of spin coherent states, or states of the Lie
group $ \mbox{SO(3)}\sim \mbox{SU(2)}$. They are parameterized by a
unit vector, the direction of the latter coincides with the quantum
expectation values for the spin operator $ {\rm {\bf m}}$, and all
quantum expectation values  of some products of spin components are
expressed through corresponding products of expectation values for
components of spin operators (dipolar variables), see for review
Refs.\onlinecite{PerelomKS,FRA91}.

For a Heisenberg ferromagnet with a purely bilinear exchange ($
K=0)$ the approach based on $ \mathrm{SO(3)}-$coherent states is
exact, whereas for the model (\ref{eq1}) one has to take into
consideration quantum expectation values for all irreducible
operators, which include not only dipolar variables $ {\rm {\bf
m}}$, but also so-called quadrupolar variables, bilinear on
components of spin operator {\rm {\bf S}}. In principle, such
variables can not be reduced to the $ {\rm {\bf m}}$ only; for
example, $\langle S_x^2-S_y^2 \rangle$ can be non-zero even the
values $\langle S_x \rangle =0$ and $\langle S_y \rangle =0$. In
fact, Hamilton dynamics of the variable $ m=\vert {\rm {\bf m}}\vert
$ takes place, and a variable, canonically conjugated to $ m$, is
quadrupolar variables of a structure mentioned above. Such dynamics,
which is appropriately called \textit{longitudinal }, in principle
does not exist for a Heisenberg ferromagnet while considered within
the framework of the Landau-Lifshitz equation or the Hamiltonian
(\ref{eq1}) with $ K=0$.

The fast change of the length of the magnetization $ {\rm {\bf m}} $
is of great interest now. Thermal quenching of magnetization length,
caused by ultrafast (femtosecond) laser pulse, is known for
different ferromagnets nearly 10 years.\cite{Bigot,Scholl,Hohlfeld}
Non-thermal laser control of magnetization is also
realized,\cite{PulseGarnets,Stanciu07} see for review
Ref.~\onlinecite{exper-rew}. Principal possibility of dynamical
(besides heating) quenching of $m=|{\rm {\bf m}}|$ till the values
$m \approx 0$ and ultrafast dynamics of the variable $m$ is of great
scientific and technological interest.

For the model (\ref{eq1}) 1D solitons\cite{IvKhymynJETP07} and 2D
topological solitons in the collinear nematic
phase,\cite{IvanovKol03,IvanovPZh06} and near the $
\mbox{SU}(\ref{eq3})-$ symmetrical point,\cite{MikushMoskvin} have
been studied. However, non-linear dynamics for other phases, even
for the simplest ferromagnet phase, has not been studied yet. In
this work we investigate 2D longitudinal non-linear spin
oscillations and solitons in the ferromagnetic phase of a
non-Heisenberg ferromagnet within a consistent semiclassical
description of the model (\ref{eq1}). In the Section 2 the equations
for a full set of spin quantum expectation values obtained within
semiclassical approximation and describing effects of dynamical
quantum spin reduction are discussed. Non-linear ``longitudinal''
spin oscillations in such the system are found there. Interaction of
corresponding non-dipolar degrees of freedom with electro-magnetic
field is also discussed in this Section.

In next Sections we have demonstrated the presence of specific
\emph{longitudinal solitons}, for which direction of the
magnetization vector $\mathbf{m}$ remains constant, however, the
magnetization changes in length. Such soliton solutions are obtained
in the framework of the semiclassical equations, in a continual
approximation (Section 3) and by analyzing of a discrete problem for
a simple square lattice (Section 4). The Section 5 contains
conclusions and discussions of results obtained, as well as some
overview of open problem. The discussion of the possibility of
excitation of longitudinal spin oscillations by ultrashort  laser
pulse are also present in this concluding section.

\section{Model and elementary excitations.}
To develop a semiclassical theory describing a magnet with a spin $
S=1$ with Hamiltonian \eqref{eq1} and to make allowance for the spin
reduction on a lattice site, we introduce generalized coherent
states of $ \mbox{SU(3)}$ group parameterized by a three-dimensional
complex vector $ {\rm {\bf u}}+i{\rm {\bf v}}$, see
Refs.~\onlinecite{IvanovKol03,MikushMoskvin},
\begin{equation}
\label{eq2} \left| {{\rm {\bf u,v}}} \right\rangle
=\sum\limits_{j=x,y,z} {\left( {u_j +iv_j } \right)\;\left| {t_j }
\right\rangle } \ ,
\end{equation}
where $ \vert t_j \rangle$ are three Cartesian states for spin $
S=1$, $ {\rm {\bf u}}$ and $ {\rm {\bf v}}$ are real vectors. With
account taken of the normalization requirements and an arbitrarity
of the total phase, the vectors $ {\rm {\bf u}}$ and $ {\rm {\bf
v}}$ satisfy the conditions
\begin{equation}
\label{eq3}
{\rm {\bf u}}^2+{\rm {\bf v}}^2=1,{\rm {\bf uv}}=0.
\end{equation}
In terms of the variables $ {\rm {\bf u}}$ and $ {\rm {\bf v}}$, all
irreducible quantum expectation values for spin $ S=1$ states,
including the magnetization vector $ \langle{\rm {\bf
S}}\rangle={\rm {\bf m}}$ and quadrupolar variables $ \langle S_i
S_j+S_jS_i \rangle$, can be expressed by the simple relations
\begin{equation}
\label{eq4} {\rm {\bf m}}=2({\rm {\bf u}}\times {\rm {\bf
v}}),\langle S_i S_k +S_k S_i \rangle =2(\delta _{ik} -u_i u_k -v_i
v_k ).
\end{equation}
For the ferromagnetic ground state, which is stable for $ J>K$, $
J>0$, the value of $ \left| {\rm {\bf m}} \right|=1$, while the
state is degenerated in the direction of $ {\rm {\bf m}}$. It means
that in the ground state $ \vert {\rm {\bf u}}\vert =1/\sqrt 2 $ è $
\vert {\rm {\bf v}}\vert =1/\sqrt 2 $. At $ \left| {\rm {\bf m}}
\right|=1$ rotation of these vectors in plane perpendicular $ {\rm
{\bf m}}$, does not change the state of a system. However, for any $
\left| {\rm {\bf m}} \right|<1$, states, which differ from each
other by direction of $ {\rm {\bf u}}$ and $ {\rm {\bf v}}$ in the
plane, are physically distinguished due to anisotropy of quadrupolar
variables. As we will see below, the angle of rotation of $ {\rm
{\bf u}}$ and $ {\rm {\bf v}}$ plays a role of a generalized
coordinate conjugated to the magnetization length $m = {\rm {\bf
m}}$.

Dynamics of the variables $ {\rm {\bf u}}$ and $ {\rm {\bf v}}$ for
given spin on a point $i$ in a lattice are determined by
Lagrangian,\cite{IvanovKol03}
\begin{equation}
\label{eq5} L=-2\hbar \sum\limits_ i {\rm {\bf v}}_i \cdot (\partial
{\rm {\bf u}}_i /\partial t)-W\{{\rm {\bf u}},{\rm {\bf v}}\}.
\end{equation}
where $ W\{{\rm {\bf u}},{\rm {\bf v}}\}$ is the system energy,
which coincides with the quantum expectation value of the
Hamiltonian (\ref{eq1}) calculated with the coherent states
(\ref{eq2}). For a lattice discrete model, an expression for the
energy $ W\{{\rm {\bf u}},{\rm {\bf v}}\}$ is given in
Ref.~\onlinecite{IvanovKol03}. Based on this Lagrangian, we can
easily analyze both the linear and nonlinear dynamics of the
ferromagnet. In particular, using the explicit form of the energy $
W\{{\rm {\bf u}},{\rm {\bf v}}\}$ proposed in,\cite{IvanovKol03} we
can readily obtain the spectrum of linear elementary excitations
(magnons). This spectrum contains two modes. The first mode does not
depends on the biquadratic interaction constant $ K$. Its dispersion
relation has the same form as for usual Heisenberg ferromagnet, $
\varepsilon ({\rm {\bf k}})=4J\left( {1-C\left( {\rm {\bf k}}
\right)} \right)$, where $ C\left( {\rm {\bf k}} \right)=(1/2)[\cos
(k_x a)+\cos (k_y a)]$, ${\rm {\bf k}}={\rm {\bf p}}/ \hbar $, ${\rm
{\bf p}}$ is the magnon momentum, $ a$ is the lattice constant
(below we will limit ourselves with two-dimensional square lattice).
In the long wave limit $ ka \ll 1,\;k=\vert {\rm {\bf k}}\vert $ the
usual parabolic dispersion law appears, $ \varepsilon  \simeq
J(ka)^2$. The second mode describes the oscillations of the modulus
of magnetization, $ m=|{\rm {\bf m}}|$, coupled with some
quadrupolar variables. It is natural to call them
\textit{longitudinal magnons}, see below Eqns. (\ref{eq666},
\ref{eq9}).

It is difficult to analyze nonlinear dynamics of the variables $
{\rm {\bf u}}$ and $ {\rm {\bf v}}$, since we have to operate with
four independent nonlinear equations rather than with two equations
for angular variables, as in the case of the usual ferromagnet.
However, it is possible to show that the full set of nonlinear
equations for the $ {\rm {\bf u}}$ and $ {\rm {\bf v}}$ vectors has
a partial planar solution for which the magnetization vector changes
its length only, $ {\rm {\bf m}}=m{\rm {\bf e}}_3 $ and vectors $
{\rm {\bf u}}$ and $ {\rm {\bf v}}$ rotate in the perpendicular
plane $ (1,\ 2 )$, where $ {\rm {\bf e}}_1 $ and $ {\rm {\bf e}}_2 $
are unit vectors in the plane, perpendicular to the magnetization
vector $ {\rm {\bf m}}=m{\rm {\bf e}}_3$, $ {\rm {\bf e}}_i $,
$i=1,2,3 $ present an orthogonal set of unit vectors. Below, we will
restrict ourselves to the analysis of such planar solutions. For
this solution, only three quantum expectation values are
non-trivial, namely, the magnetization $ \sigma _3 =m=2(u_1v_2 -u_2
v_1 )$ and two quadrupolar variables, $ \sigma _1 =\langle S_1^2
-S_2^2 \rangle =u_1^2 -u_2^2 +v_1^2 -v_2^2 $ and $ \sigma _2
=\langle S_1 S_2 +S_2 S_1\rangle =-2u_1 u_2 -2v_1 v_2 $. One can
easily show that $ \sigma _1^2 +\sigma _2^2 +\sigma _3^2 =1$; that
is, the vector $ \boldsymbol{\sigma} =\sigma _1 {\rm {\bf e}}_1
+\sigma _2 {\rm {\bf e}}_2 +\sigma _3 {\rm {\bf e}}_3 $ is a unit
vector. It is convenient to introduce the angular representation for
this unit vector,
\begin{eqnarray}\label{spinvars}
\nonumber \sigma _3 &=& \cos \theta, \ \sigma _1=\langle
S^2_x-S^2_x\rangle
= \sin \theta \cos \phi,\\
\sigma _2 &=&\langle S_xS_y+ S_yS_x\rangle = \sin \theta \sin \phi .
\end{eqnarray}

The advantage of these variables is that they are unambiguously
determined from a given physical state of the system. In contrast,
the variables $ {\rm {\bf u}}$ and $ {\rm {\bf v}}$ contain the
halved values of the angular variables $ \theta ,\;\phi $, that
reflects the nature of vector $ {\rm {\bf u}}$ (or $ {\rm {\bf v}})$
as a vector --- director. Using the angular variables $ \theta ,
\phi $, we can reduce  the Lagrangian (\ref{eq5}) to the form
\begin{equation}
\label{discrete} \mathcal{L}=\frac{\hbar}{2}\sum _i(\cos \theta _i
-1)\cdot \bigl(\frac{\partial \phi _i } {\partial t}\bigr)
-W\{\theta ,\phi \},
\end{equation}
where $ W\{\theta ,\phi \}$ is the system energy, which depends on
the discrete variables $ \theta _i$, $\phi _i $. It is convenient to
present the energy through the vector variable
$\boldsymbol{\sigma}$,
\begin{equation}
\label{XXZ} W=-\frac{1}{2}\sum_{<i,j>}
\left[K\boldsymbol{\sigma}_{i} \boldsymbol{\sigma}_{j} + 2(J-K)
\sigma_{i,3} \sigma_{j,3} \right].
\end{equation}
It is interesting to note, this Lagrangian formally coincides with
that for a spin $ S=1/2$ uniaxial ferromagnet which is known as $
\mbox{XXZ}$ --model. For this model, the constant of isotropic
exchange equals to $ K/2$ and anisotropy of spin interaction is
proportional to $ J-K$. Thus, the general dynamics of $
\mbox{SU(3)}-$ coherent state for spin one magnet includes the
particular class of solutions, which are described by
\textit{classical }model for a spin $ S=1/2$, that is quite unusual
model for theory of magnetism. Limit cases correspond to the
following simple physical models: in the vicinity of a transition to
the nematic phase, $ \mbox{SU(3)}-$ symmetrical point $ K\to J $, an
effective spin model (\ref{XXZ}) becomes isotropic, while at $
K/J\to 0$ we arrive at the Ising model. Naturally, anisotropy of the
effective model is realized in the $ \boldsymbol{\sigma }$ space and
has not direct linkage to spacious rotations of spin operators.

For the model (\ref{discrete}) it is easy to obtain oscillations,
which are a non-linear analogy of the above mentioned longitudinal
magnons. For this excited states, at the lattice site $\mathbf{l}_i$
the variable $ \theta _i =\theta _0 =\mathrm{const}$ not depends on
time and $ \phi _i ={\rm {\bf kl}}_i +\omega t$. The frequency of
such oscillations $ \omega $ depends on the wave vector $ {\rm {\bf
k}}$ and the amplitude $ \theta _0 $ as following
\begin{equation}
\label{eq666} \omega ({\rm {\bf k}})=(4\cos \theta _0 /\hbar
)\bigl\{2(J-K)+K[1-C({\rm {\bf k}})] \bigl \}.
\end{equation}

Within the linear approximation, at $ \theta _0 \to 0$, this
frequency becomes the frequency of longitudinal magnons, previously
obtained by Papanicolaou.\cite{PapanicoFM} In the long-wave limit
this spectrum becomes parabolic, and can be written as
\begin{equation}
\label{eq9}
\omega ({\rm {\bf k}})=\omega _0 \cos \theta _0 (1+r_0^2 k^2),
\end{equation}
where
\begin{equation}
\label{eq10} \hbar \omega _0 =8(J-K),\;r_0^2 =Ka^2/8(J-K).
\end{equation}

Here the frequency $\omega _0 $ is a gap of longitudinal magnons, $
r_0$ determines a characteristic space scale. A study based on the
Lighthill criterion, see for example Ref.~\onlinecite{Whitham},
shows that such uniform oscillations being excited in the system
(for example, by ultrafast laser pulse, see below) are unstable
against self-focusing. As a result, essentially non-uniform states,
like solitons, should appear. For their analysis it is easier to
employ a continual approximation, considering $ \theta _i , \phi _i
$ to be continuous functions of coordinates and time, $ \theta _i
(t)\to \theta  ({\rm {\bf x}},t),\phi _i (t)\to \phi ({\rm {\bf
x}},t)$. For 2D system, or for a thin enough film of magnet, which
complies with standard geometry of experiment,\cite{exper-rew} one
can use 2D solutions, and present the Lagrangian of the problem as $
\mathcal{L}= \int L\{\theta ,\phi \}\cdot d^2x,$ where the density
of Lagrangian is
\begin{equation}
\label{eq11} L=(\hbar /2a^2)(\cos \theta -1)(\partial \phi /
\partial t)-w\{\theta ,\phi \},
\end{equation}
the energy density $ w\{\theta ,\phi \}$ is determined by the
formula
\begin{equation}
\label{eq12} {\begin{array}{*{20}c}
 {w\{\theta ,\phi \}=(2/a^2)(J-K)\sin ^2\theta +(1/4)K\sin ^2\theta \mathop
{\left( {\nabla \phi } \right)}\nolimits^2 \,+} \hfill \\
 {+(1/4)\left[ {K+2(J-K)\sin ^2\theta } \right]\mathop {\left( {\nabla
\theta } \right)}\nolimits^2 \,.} \hfill \\
\end{array} }
\end{equation}
As we will see below, solitons exist at $0< K < J$, and we will
limit ourselves to this region of the parameters.

Let us discuss briefly an interaction of longitudinal degrees of
freedom with external fields, having in mind primarily a possibility
of experimental excitation of such oscillations. First of all, a
magnetic field $ {\rm {\bf H}}=H{\rm {\bf e}}$ affects only the
magnetization, for a planar solution it is $ {\rm {\bf
m}}=\langle{\rm {\bf S}}\rangle={\rm {\bf e}}_3 \cdot \cos \theta $.
Therefore Zeeman interaction of $ \boldsymbol{\sigma }$ with the
magnetic field parallel to some direction $ {\rm {\bf e}}$ is
described by the Hamiltonian $ H^H=-H({\rm {\bf e}},{\rm {\bf e}}_3
)\sigma _3 $. Actually, the magnetic field directed in parallel with
the mean spin $ {\rm {\bf m}}=\langle{\rm {\bf S}}\rangle$ does not
affect the system state, while for any other directions of $ {\rm
{\bf H}} $ one can expect trivial change of the orientation of unit
vectors $\mathbf{e}_i$ describing the planar solution. It turns out,
that ac- magnetic field is not effective for excitation of
longitudinal oscillations.

It appears, that excitation of longitudinal oscillations may be done
through application of an electric field on the system.  For
simplicity, we will start with dc--electric field. Interaction of
such electric field with a spin system of a magnet can be described
phenomenologically on the basis of the following Hamiltonian
\begin{equation}
\label{eq13} H^{(int)}=\frac{1}{8\pi }\varepsilon
_{ij}^{\mathrm{(spin)}} E_i E_j ,
\end{equation}
where $ \varepsilon _{ij}^{\mathrm{(spin)}} $ is a spin-dependent
part of dielectric permittivity and $ {\rm {\bf E}}=E{\rm {\bf e}}$
is the electric field, ${\rm {\bf e}}^2=1$, see
Ref.~\onlinecite{LL-SploshSredy}. In principle, $ \varepsilon
_{ij}^{\mathrm{(spin)}} $ can include all spin variables, describing
the system state and allowed by symmetry.\cite{Turov+,Zvezdin} In
our case, the components of $ \varepsilon _{ij}^{\mathrm{(spin)}} $
should include the contribution from quadrupolar variables, $
\varepsilon _{ij}^{\mathrm{(spin)}} E_i E_j =\varepsilon ^Q \cdot
\langle S_i S_j +S_j S_i \rangle E_i E_j $. The discussion of the
microscopic origin of such interaction, in particular, the value of
the constant $\varepsilon ^Q$, is far from the scope of this
article.

A possible role of quadrupolar interactions can be demonstrated by
simple example. Consider an electric field, perpendicular to the
magnetization in the ground state ${\rm {\bf m}}=m {\rm {\bf e}}_3
$. It can be easily seen, the influence of such field is equivalent
to the action of some ``effective field'' $ {\rm {\bf H}}^Q$ on the
variable $\boldsymbol{\sigma }$ of the same form as the Zeeman
interaction of usual magnetic field with usual magnetic moment, $
H^{\mathrm{(int)}}=-{\rm {\bf H}}^Q\boldsymbol{\sigma }$. Here the
effective field $ {\rm {\bf H}}^Q$ is described by
\begin{multline} \label{eq14}
\mathbf{H}^Q = \frac{\varepsilon ^Q}{8\pi }
\{[(\mathbf{e}_1,\mathbf{E})^2 -
(\mathbf{e}_2,\mathbf{E})^2]\mathbf{e}_1+
2(\mathbf{e}_1,\mathbf{E})(\mathbf{e}_2,\mathbf{E})\mathbf{e}_2\} =
\\ = \frac{\varepsilon^QE^2}{8\pi }
(\mathbf{e}_1\cos \alpha+\mathbf{e}_2\sin \alpha ),
\end{multline}
where $ \alpha $ is a doubled value of the angle between the vector
of electric field $ {\rm {\bf E}}$ and the direction of unit vector
$ {\rm {\bf e}}_1 $ in planar solution. Then a simple analogy
between action of the usual magnetic field $ \mathbf{H}$ on the
magnetization $\textbf{m}$ and action of the field $ \mathbf{H}^Q$
on the vector $\boldsymbol{\sigma }$ becomes obvious.

For consideration of ac-field, for example, electric field of
electromagnetic wave (light) it is enough to replace $ (1/8\pi )E_i
E_j $ by $ (1/16\pi )E_i E_j^\ast $, where $ E_i $ now is the
complex amplitude of the time-dependent electric
field.\cite{LL-SploshSredy} For linearly polarized light, the same
expression \eqref{eq14} for the effective field appears. If light
intensity is time-dependent, for example, for modulated laser beam
or for ultrashort laser pulse, the effective field $\textbf{H}^Q$
will be time-dependent. Being linearly coupled with variables
$\sigma _1, \sigma _2$, it can excite the longitudinal spin
oscillations found above.

\section{ Longitudinal solitons in a continual model of a magnet. }
As Lagrangian does not depend directly on $ \phi $, but on its
derivatives, the model (\ref{eq11},\ref{eq12}) has an integral of
motion, which determines the total spin projection on some axis. The
same  integral of motion is present for discrete model
(\ref{discrete},\ref{XXZ}). This integral of motion is suitable to
be presented via the number of spin deviations in the system $ N$,
for discrete model and its continuous counterpart it reads
\begin{equation}
\label{eq15} N=\sum\limits_i (1-\cos\theta_i) , \ \mathrm{or} \ N=
\int{(1-\cos\theta)d^2x/a^2}.
\end{equation}
In the framework of quantum mechanics, $ N$ possess integer values
and in line with Ref.~\onlinecite{Kosevich+All} we use this quantity
for semiclassical quantization of solitons. The important integral
of motion is soliton energy $ E$ (\ref{XXZ}) which within the
continuous approximation turns into $ E=\int w\{\theta ,\phi \}
d^2x$.

The integral of motion \eqref{eq15} results in solutions with
stationary dynamics, for which the vectors $ {\rm {\bf u}}$ and $
{\rm {\bf v}}$, as well as in-plane components of the vector
$\boldsymbol{\sigma }$ rotate in the plane $ (1,2)$ with some
constant frequency.  In this case, the variable $ \theta $ depends
only on a distance from a certain point in the plane, which is
considered as the center of a soliton. We limit ourselves to
analysis of these solutions basing on ansatz of the form
\begin{equation}
\label{eq16} \theta =\theta (r),\varphi =\omega t,\;r=\vert {\rm
{\bf x}}\vert ,
\end{equation}
Such solitons formally resemble so-called precessional solitons,
which are known for the case of uniaxial Heisenberg ferromagnets
with precession of a unit vector of magnetization $ {\rm {\bf m}}$,
$ \vert {\rm {\bf m}}\vert =1$, around an easy axis (z-axis) with a
constant frequency $ \omega _H $ and with an amplitude $ m_{\bot}
=\sin \vartheta (r)$ depending on $ r$, $ \vartheta (r)\to 0$ at $
r\to \infty $, see for review Ref.~\onlinecite{Kosevich+All}. In
spite of principal difference in physical properties of these two
types of solitons many formal features of them are similar. The
latter allows us not to discuss some details.

The function $ \theta (r)$ is determined by a ordinary differential
equation,
\begin{eqnarray}
\label{eq17} \nonumber ( \theta ^{\prime \prime} &+&\frac{\theta
^{\prime}}{r} )\cdot \left( r_0^2 +\frac{a^2}{4}\sin ^2\theta
\right)+
\\&+& \frac{\omega }{\omega _0}\cdot \sin \theta -\sin \theta \cos
\theta \left( 1-\frac{\theta ^{\prime 2} a^2}{4} \right)=0,
\end{eqnarray}
where the prime denotes the derivative over $ r$, the characteristic
size $ r_0 $ and the magnon gap frequency $ \omega _0 $ are
determined above (\ref{eq10}). Far from a soliton the state should
correspond to the ground state of the system, i.e. the condition $
\theta (r)\to 0$ at $ r\to \infty $ should be fulfilled. The
condition $ {\theta }'(0)=0$ ensures the absence of singularity at $
r=0$. The equation (\ref{eq17}) with such boundary conditions can be
easily solved by the ``shooting'' method.\cite{Kosevich+All} It has
a discrete set of solutions $ \theta =\theta _n (r)$ with $ n$ nodes
at points $r= r_n \ne 0$, $n=\;1,\;2,...$. Solitons with nodes are
unstable,\cite{Kosevich+All} and we will discuss the solution with $
n=0$ and a monotonous decay of the function $ \theta (r)$ only.

Knowing the solution of $ \theta (r)$ one can calculate integrals
describing $ E$ and $ N$, and represent the soliton energy as a
function of the number of magnons bounded in the soliton, $E=E(N)$.
As already mentioned this procedure at $ N\gg 1$ and within the
continual approximation is equivalent to semiclassical quantization
of solitons (for the discrete model some peculiarities occur, see
Section 5). It is convenient to use the fact that the equation
(\ref{eq17}) can be formally obtained from variation of the
functional $ \tilde {\mathcal{L}}=E-\hbar \omega N$, $\delta \tilde
{\mathcal{L}}/ \delta \theta =0$. The functional $ \tilde
{\mathcal{L}}$ coincides with Lagrangian (\ref{eq5}) calculated
within the ansatz (\ref{eq16}). The condition $\delta \tilde
{\mathcal{L}}/\delta \theta$ immediately leads to the relation
\begin{equation}
\label{eq18} \hbar \omega =dE(N)/dN,
\end{equation}
which coincides with that for precessional solitons. Eq.
(\ref{eq18}) describes the quantum sense of the classical parameter
$ \omega $ in the solution of the form (\ref{eq16}): the value of $
\hbar \omega $ at $N \to \infty$ equals to a change of the soliton
energy with a change of the number of bound magnons by one.

Some limit characteristics of solitons can be obtained without an
exact solution of (\ref{eq17}). Using the phase plane method, it is
easy to demonstrate that a soliton solution exists only at $ 0<
\omega  < \omega _0$ and its characteristics depend on the parameter
$ \omega /\omega _0 $. At $ (1-\omega /\omega _0 ) \ll 1$ the
soliton amplitude is small and the function $\theta (r)$ takes the
form
\begin{equation}
\label{eq19}
\theta (r)=\sqrt {1-\omega /\omega _0 } \cdot f\left[ {(r/r_0 )\sqrt
{1-\omega /\omega _0 } } \right],
\end{equation}
where $ f(x)$ is a universal function, localized in the area of $
\Delta x\sim 1$ with the value of the order of one in the origin.
Further it is possible to demonstrate that for all $ K/J$ at $
\omega \to \omega _0 $ soliton energy tends to a finite value $ E\to
E_0 =\eta \cdot 4\pi K$, where $ \eta \approx 0.93$ is a numerical
coefficient. In this limit the number of magnons is also finite, $
N\to N_{\min } =2\eta N_2 $, where $ N_2 =2\pi (r_0 /a)^2$ is the
characteristic number of magnons.

These values are minimal for solitons in a magnet with a given
parameter $ K/J$, and in the limit $ \omega \to \omega _0 $ the
connection $ E\to \hbar \omega _0 N$, which is typical for linear
theory, appears. The similar property takes place for a precessional
soliton with small amplitude,\cite{IvZaspYastr} with an essential
difference, that for a precessional soliton the value of $ E_0 $ is
always of the order of the exchange integral $ J$ and can be
compared with energy of a Belavin-Polyakov topological soliton $
E_{BP} =4\pi J$, whereas for a longitudinal soliton for small $ K/
J$ the inequality $ E_0 \ll E_{BP} $ formally may be realized. In
fact, for $ K \ll J$ the continual approximation is failed even at
$N \sim 1$, and the minimal value of soliton energy $E$ can not be
smaller then $2.57J$, see the last paragraph of the next Section. On
the other hand, the value of $ E_0 $ found here within the continual
approximation is valid for wide region of parameters like
$0.7J<K<J$, where the value of $N_{\min }$ is larger then 4, see the
next Section.

Another limit case corresponds the condition $ 0<\omega \ll \omega
_0 $. To discuss it, we can mention that for $ \omega =0$  the 2D
soliton solution is absent, however, equations allow a 1D
``longitudinal'' domain wall-like solution. For this solution, $
{\rm {\bf m}}(\xi )=m(\xi ){\rm {\bf e}}_3 $, with $ m(\xi )\to 1\ $
at $ \xi \to \infty $ and $ m(\xi )\to -1$ at $ \xi \to -\infty $, $
\xi$ is a coordinate along some direction in the magnet's plane.
This wall has a characteristic width of the order of $ r_0 $ and
energy $ \sigma =\sigma (K,J)$ per unit length. A qualitative
analysis of (\ref{eq17}) gives that at $ 0<\omega \ll \omega _0 $ a
soliton contains a large enough circular region with a radius $ R
\gg r_0 $ separated form the rest of the magnet by such a wall. Here
again the situation is common to that for precessional
solitons.\cite{Kosevich+All}

Further it is easy to obtain a qualitative description of a soliton
in this limit case. Apparently, that a uniform state with $ m(\xi
)=-1$ has the same energy as at $ m(\xi )=1$, and a finite region
with $ m(\xi )=-1$, i.e. $ \theta =\pi $, does not contribute to
energy, but affects the value of $ N$. In this case energy loss is
connected only with the presence of a domain wall separating the
inner region from the rest of the magnet. For the circular area
energy loss is minimal at given $ N$ and one can see that $ N=2\pi
R^2$ and $ E=2\pi \sigma R$, where $ R$ is the radius of this area.
Proceeding from that one can obtain a square root dependence of
soliton energy  on the number of bounded magnons $ N$ for a large
soliton radius, which corresponds to the condition $ N\gg N_2$,
\begin{equation}
\label{eq20} E=a\sigma \sqrt {2\pi N}
\end{equation}

\begin{figure}[!tb]
\includegraphics[width=\figwidth]{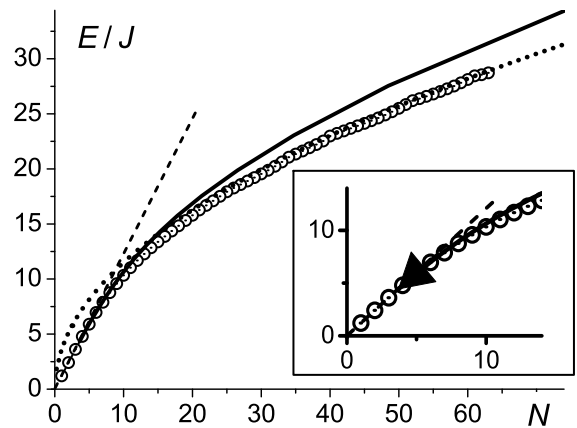}
\caption{The soliton energy $ E$ (in the unity of the exchange
integral $ J$ ) dependence on the number of magnons $ N$ for the
case $ K=0.7J$. The results of the continual approximation are
represented by the solid line, the symbols are results of a
numerical analysis of the discrete model, see the next Section. In
the inset the detail behavior at small $ N$ is present; the arrow's
end points out the end point of the dependence $ E(N)$, i.e. minimal
value of soliton energy. The symbols beneath the arrow depict
delocalized magnon states found by numerical simulation for a finite
system.  The dash line and doted line  represent asymptotic
behaviors at $ E\to \hbar \omega _0 N$ and square root asymptotic
(\ref{eq18}) for small and large $N$, respectively.} \label{fig1}
\end{figure}

\begin{figure}[!tb]
\includegraphics[width=\figwidth]{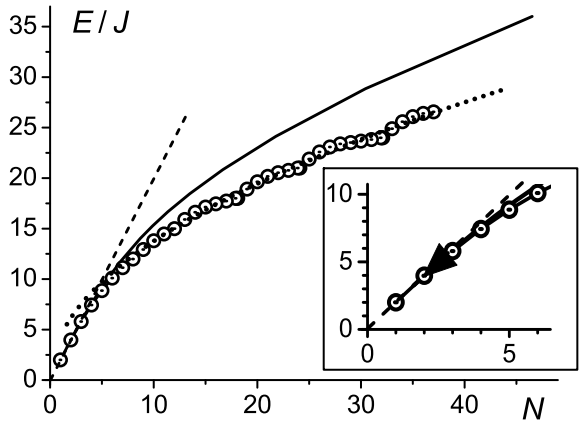} \label{fig2}
\caption{ The same as in Fig.1, but for the case $ K=0.5J$. Please
note essentially smaller values of $ N_{\min }$ and $ E_0 $}
\end{figure}

Thus in the limit cases the dependence of soliton energy on $ N$,
and also the parameters $ J,\;K $ is easily reconstructed. In the
intermediate frequency range, which corresponds to numbers of
magnons $ N$ of the order of few $ N_{\min } $ a thorough analysis
of (\ref{eq17}) is needed.  We carried numerical calculations for a
set of values of $ K/J$ for the region of interest $0<K<J$. The
analysis was done as follows: at the given $ K/J$ the equation
(\ref{eq17}) was solved numerically for a set of values of $ \omega
/\omega _0 $, which were chosen with different steps. Further having
a solution the value of energy $ E$ and the number of magnons $ N$
were calculated. Then the dependencies $ E(N)$ and $ \omega (N)$
(the latter is important for analysis of stability of a soliton,
which is stable in continual model at $d \omega (N)/dN >0$ only)
were constructed.

Let us briefly discuss the main characteristics of solitons found
within continual approximation. The analysis confirmed asymptotic
dependencies derived above, see Figs. 1,2. In the whole region of
parameters of the problem the function $ \omega (N)$ monotonously
decreases with the $ N$ growth, i.e. the stability condition is
fulfilled. Thus, in the framework of the continual approximation
stable soliton solutions exist within the parameter region $0< K <J$
of Hamiltonian (\ref{eq1}). The soliton energies have a lower limit,
$E_0$, which is smaller then the energy of familiar Belavin-Polyakov
solitons.

\section{The discreteness effects for longitudinal solitons.}
Strictly speaking, the continual approximation is valid only when
the characteristic size of a soliton is essentially bigger than the
interatomic distance, $ \vert \nabla \theta \vert \ll 1/a$. This
condition may be met for solitons with small amplitude, see
(\ref{eq19}), and also in the limit case $ J-K \ll K$, when the
characteristic size $ r_0 $ is larger that the lattice constant $
a$, $ r_0 \gg a$. However, in contrast to precessional solitons in
Heisenberg magnets with weak anisotropy, in which the characteristic
length is tens and hundreds of the lattice constant, in our case the
condition $ r_0 \gg a$ is much stricter. Even for enough small $
J-K=0.1\cdot J$, the value $ r_0 =1.095\cdot a$ and only slightly
exceeds the lattice constant $ a$. In the region of the parameters $
K\sim J-K\sim J$ and in the especially interesting case $ K \ll J$,
when the minimal energy of solitons is small, an applicability of
this approach is not clear and one can expect essential discreteness
effects.

Let us consider the discrete model (\ref{discrete}, \ref{XXZ}) for a
square lattice. Analysis of discrete equations for the variables $
\theta _i $ and $ \phi _i $ given for each lattice site demonstrates
existence of a solution in the form of $ \varphi _i =\omega t$ and
further it is possible to study only variables $ \theta _i $.

For analysis of solitons we employ the variation procedure proposed
and numerically realized in Ref.~\onlinecite{Iv+PRB06}. We will seek
a conditional minimum of Hamiltonian, in fact, classical energy $
W(\theta _i )$, with respect to variables $ \theta _i $, under the
condition that the number of magnons $ N=\sum\nolimits_i (1-\cos
\theta_i )$ is fixed. While seeking a minimum one can find the
precession frequency $ \omega $ from the equation $ \partial
W/\partial \theta _i =\hbar \omega \cdot \sin \theta _i $. It is
worth noting, the sign of derivative $ d\omega /dN$ in a discrete
case is not important; a soliton is stable, if the found conditional
extreme of energy is minimum.  Analysis was done for an
approximately circular fragment cut from a square lattice sized
$24\times 24$. We limit ourselves to such size as states we are
interested in are essentially localized, and the influence of
borders on them is negligible, while increase of a sample size
require a significant increase of numerical calculation time. As one
can expect at small values of $ (J-K)/K$ the behavior of the
dependencies $ E(N)$ and $ \omega (N)$ merely follows curves
obtained within the continual approximation, therefore we do not
present them. As well, for a region of small $ N$ our analysis
demonstrates that even in the case $ r_0 \le a$ the results of the
continual approximation are quite close to numerical data, see. Fig.
1 and Fig. 2.

It is interesting to note, that even for large $ N$, when the
characteristic size of inhomogeneity in a solution is of the order
of $ r_0 <a$, these results qualitatively describe the dependence $
E(N)$ even at moderate values of $ K/J$, such as $ K/J=0.7$ and $
K/J=0.5$, for which $ r_0 =0.54a$ and $ r_0 =0.354a$, respectively.
For $ K/J=0.7$ numerical data adhere closely continual curves, and a
square root dependence with fitted value of domain wall energy
$\sigma$ is working rather well. Even for the smaller value $
K/J=0.5$ only insignificant sign-alternating deviations from the
square root dependence (\ref{eq20}), which are almost invisible in
Fig.~1 for $ K/J=0.7$, are observed on the numerical data on Fig.~2.

\begin{figure}[!tb]
\includegraphics[width=\figwidth]{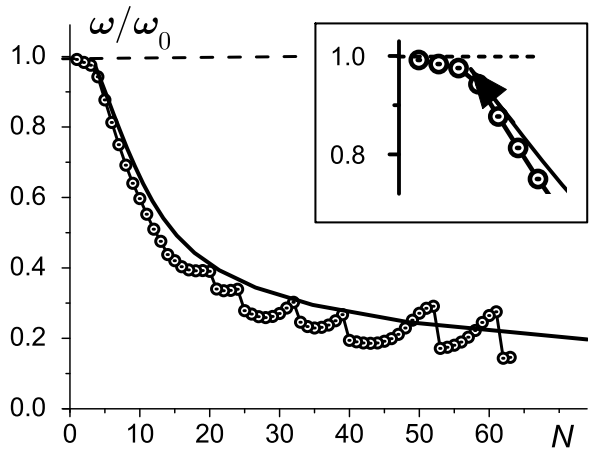}
\caption{The dependence of the oscillation frequency $ \omega $ (in
$ \omega _0 $ units) on the number of magnons $ N$ for a soliton in
the case $ K=0.7J$. The results of the continual approximation are
shown by the solid line, symbols are a result of the numerical
analysis of the discrete model. The horizontal dash line points out
theoretical value of magnon gap $\omega _0$. In the inset detail
behavior at small $ N$ is showed, the end of the arrow points out
the end point of continual dependence $N=N_{\mathrm{min}}$. Symbols
on the left side of the arrow correspond to delocalized magnon
states, their insignificant deviations from the value of $ \omega _0
$ are caused by boundary effects, naturally present for a finite
system.} \label{fig3}
\end{figure}

\begin{figure}[!tb]
\includegraphics[width=\figwidth]{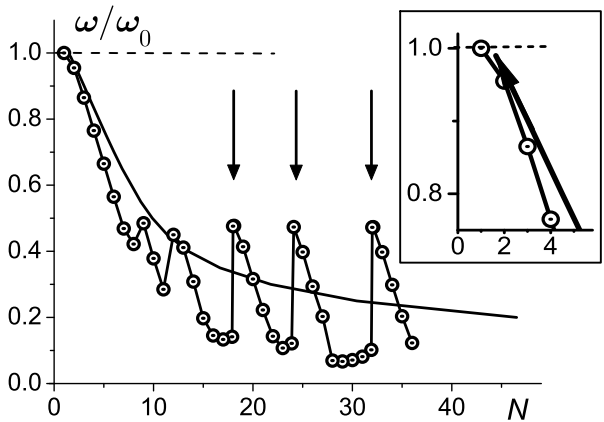}
\caption{The same as in Fig. 3, for the case $ K=0.5J$. The vertical
arrows denote special values of $ N=18,\;24$ and 32, see the text.
Please note that the dependencies correspond each other even at not
small $ (\omega -\omega _0 )/\omega _0 \le 0.25$, that corresponds $
N\le 5$. It is worth to mention that only one nonlocalized state
with $ N=1$ was revealed. } \label{fig4}
\end{figure}

However, apart from such characteristic of a soliton as $ E(N)$,
which depends on integral values $ E$ and $ N$ only, effects of
discreteness, nevertheless, are essential. This is apparently
demonstrated in the dependence $ \omega (N)$, see Fig. 3, 4 and
especially it is clearly seen in analysis of the soliton structure,
i.e. real distribution $ \theta _i $ in the lattice, see below. For
dependencies $ \omega (N)$ when $ N$ grows at the beginning, at
small $ N$ regular deviation of numerical data from the continual
curve is observed. This feature can be explained by the fact that $
\omega \propto dE/dN$, whereas the energy  $E(N)$ in the discrete
model is lower than in continuum. However, even for $ K/J=0.7$ it
was observed noticeable deviations up and down from the smooth
dependence typical for continual approximation, see. Fig. 3. For
smaller value $ K/J=0.5$, this irregular behavior is much more
essential, see Fig. 4. This complicated behavior is observed in that
region of $ N>10 \ll N{\mathrm{min}}$ where the effective domain
wall approximation and the square root asymptotic (\ref{eq20})
should be applicable. Therefore as for precessional solitons in a
Heisenberg magnet with strong single-ion anisotropy,\cite{Iv+PRB06}
it can be naturally associated with characteristics of lattice
pinning of a domain wall. Let us discuss the structure of a soliton
for $ K/J=0.5$, see Fig. 5 and 6.

\begin{figure}
 \subfigure[\ $ N$ =1]{\includegraphics[width = 41mm]{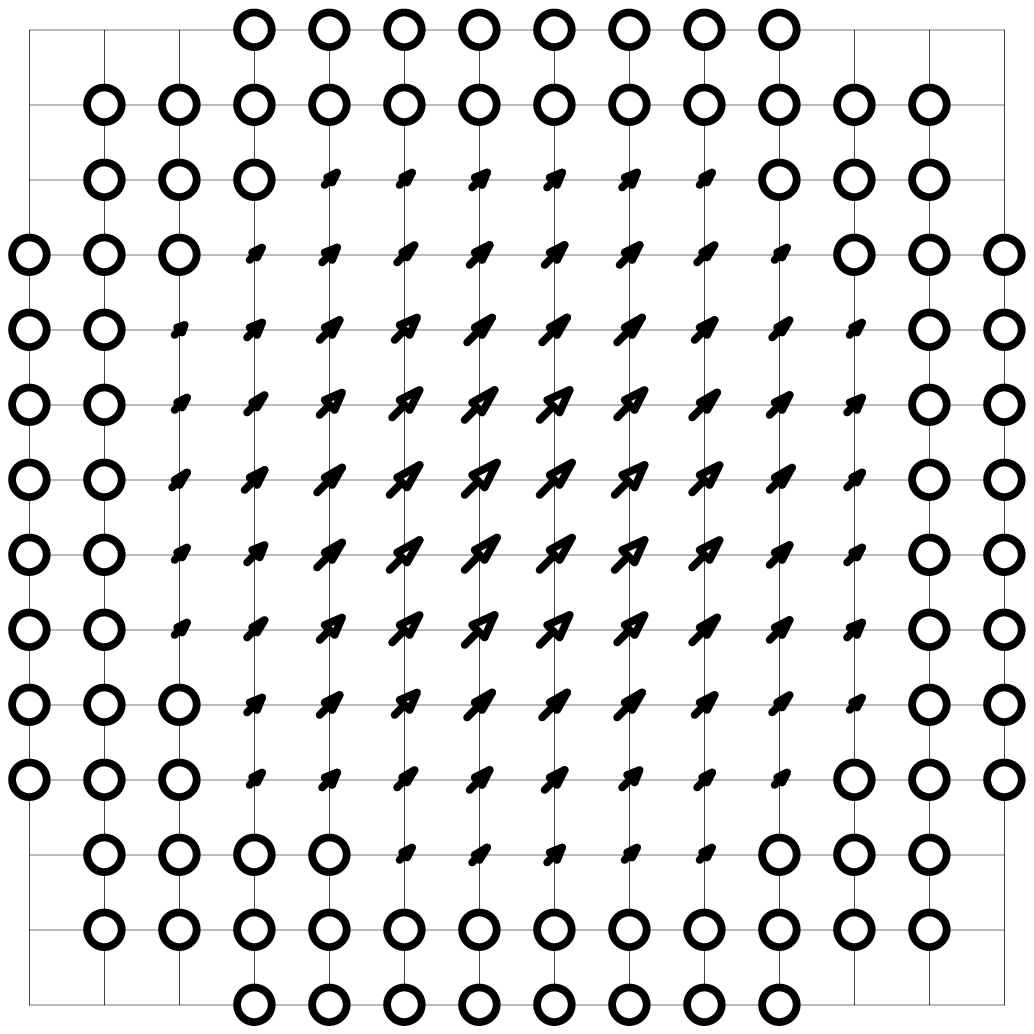}
 \label{1}}
 \subfigure[\ $ N$ =2 ]{\includegraphics[width = 41mm]{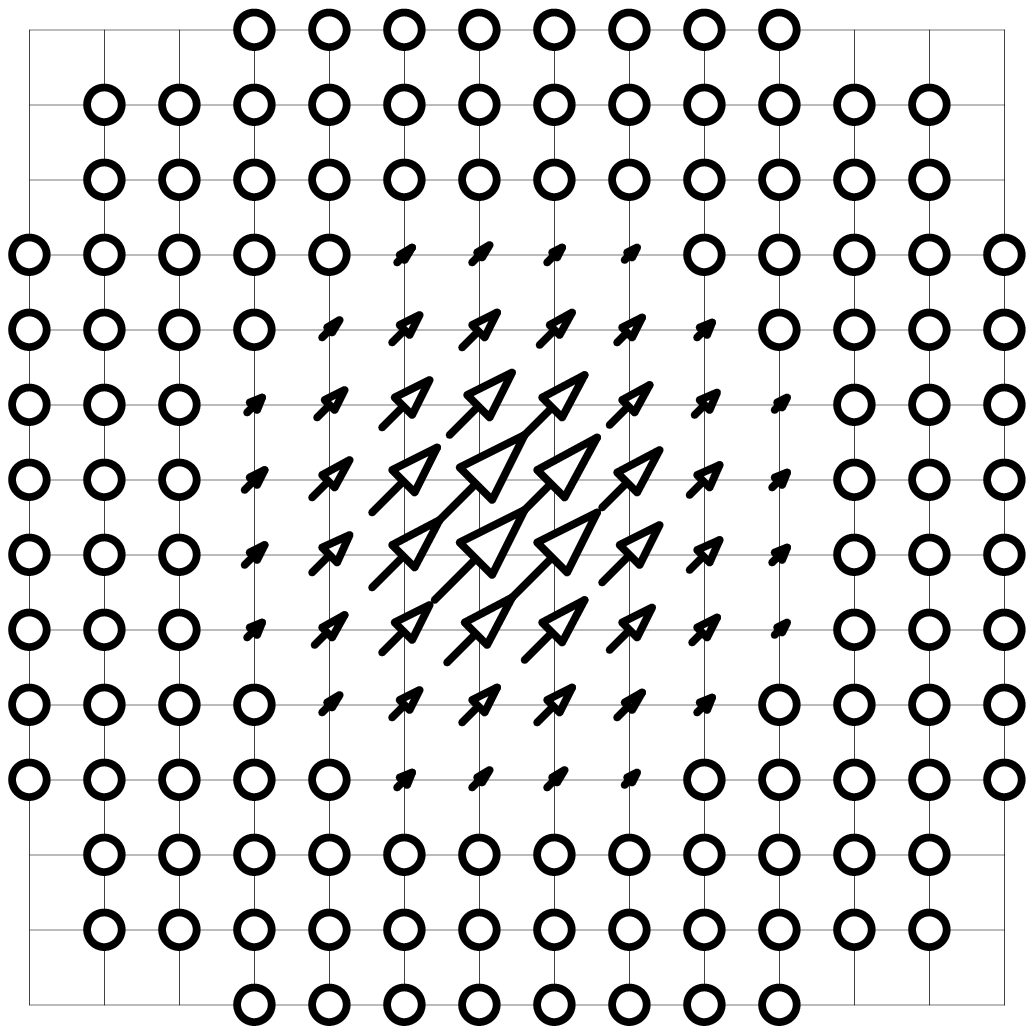}
 \label{2}}
 \subfigure[\ $ N$ =3 ]{\includegraphics[width = 41mm]{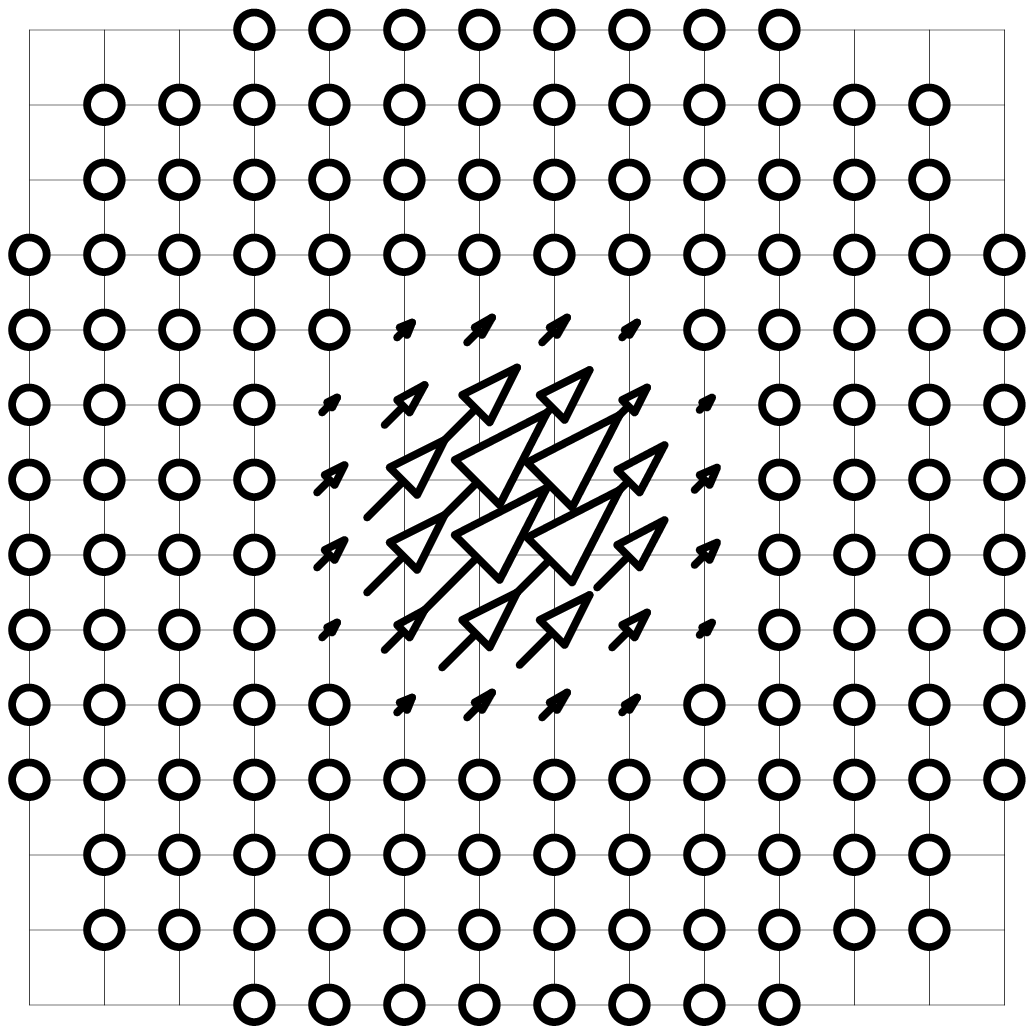}
 \label{3}}
 \subfigure[\ $ N$ =8]{\includegraphics[width = 41mm]{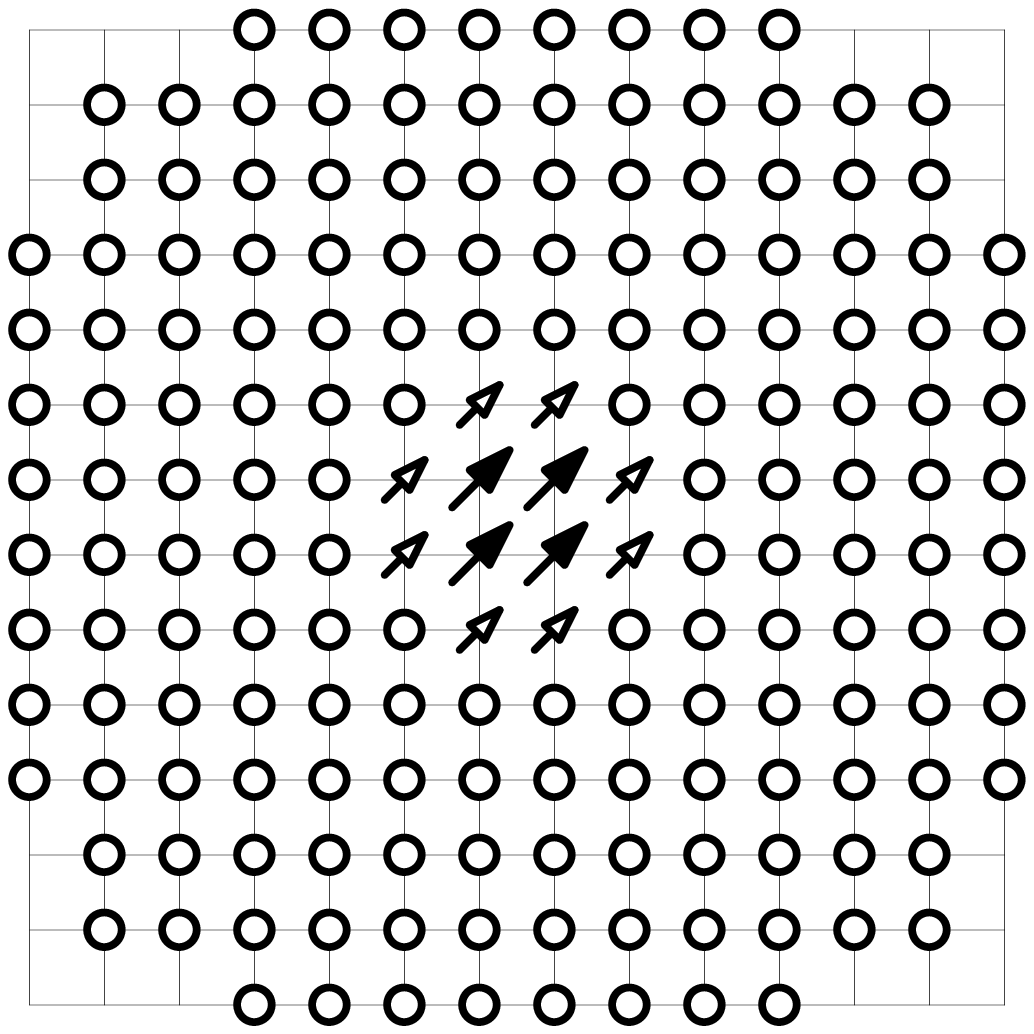}
 \label{8}}
\caption{Distribution of the discrete variables $ \theta _i $ and $
\phi _i $ in a soliton at small $ N$ for the system with $ K/J=0.5$.
Only a small part of fragment chosen for numerical calculation is
presented.  $ \theta $ and $ \phi $ for each spin are presented in
the form of 3D vector $\boldsymbol{\sigma}=(\sin \theta \cos \phi
,\;\sin \theta \sin \phi ,\;\cos \theta )$. Values of $ 5^0<\theta
_i \le 90^0$ are presented by arrows with open heads, while for $
90^0<\theta _i <180^0$ by arrows with solid heads, the small values
$0\leq \theta _i <5^0$ are presented by open circles. The transition
from non localized state at $ N=1$ to localized states at $ N>1$ is
clearly seen. For $ N= 3$ the minimal value of $ \vert {\rm {\bf
m}}\vert =\vert \cos \theta \vert \approx 0.5$ is observed in the
soliton center. For $ N=8$, the values of $\theta _i$ nearly the
soliton center are $ 158^0$ and $ 15^0$, and the tendency of
transition to collinear structure is seen. } \label{fig5}
\end{figure}

At small $ N$ numerical analysis demonstrates almost
radially-symmetric distribution of $ \theta _i $ with the scale of
few lattice constant $ a$, that is essentially larger than $ r_0
=0.354a$. It is interesting to note that even such a sensitive
parameter as minimal value of $N$ is well reproduced by the
continual calculation. According to our continuum calculation, for a
magnet with $ K/J=0.5$ the value $ N_{\min } =1.6$. The discrete
analysis provides good localization of a soliton at $ N=2,\;3$ and
much less localized state at $ N=1$, see Fig. 5. With further growth
of $ N$ up to $ N\le 10$ the distributions of $ \theta _i $ become
much more sharp and tendency of formation of collinear states is
observed. For the values  of $ N \ge 10 $ the role of discreteness
effects, primarily effects of domain wall pining, increases. This
pinning may include either the dependence of the wall energy on its
orientation related to lattice vectors or the dependence of the wall
energy on its position related to distance to corresponding atomic
lines in the lattice.

A set of our numerical results may be explained considering that,
like in the case of uniaxial Heisenberg ferromagnet, see
Ref.~\onlinecite{Iv+PRB06}, the most favorable position for a domain
wall is to be placed between atomic lines like (0,1) or (1,0), but
in contrast to Ref.~\onlinecite{Iv+PRB06}, the domain wall is quite
flexible and its bend from the line (0,1) to (1,0) does not costs
too much energy. In principle, it corresponds to conclusions of
Gochev,\cite{Gochev} who has demonstrated that for a 2D discrete
classical spin model with anisotropy like (\ref{XXZ}) pinning
effects are present for a domain wall parallel to the axis (0,1) and
(1,0) and are absent for a wall parallel to atomic lines like (1,1).
On the basis of these assumptions one can describe real distribution
of the $ \theta _i $ amplitude and complicated behavior of $ \omega
(N)$ in a lattice soliton.

\begin{figure}
 \subfigure[\ $ N$ =17]{\includegraphics[width = 27mm]{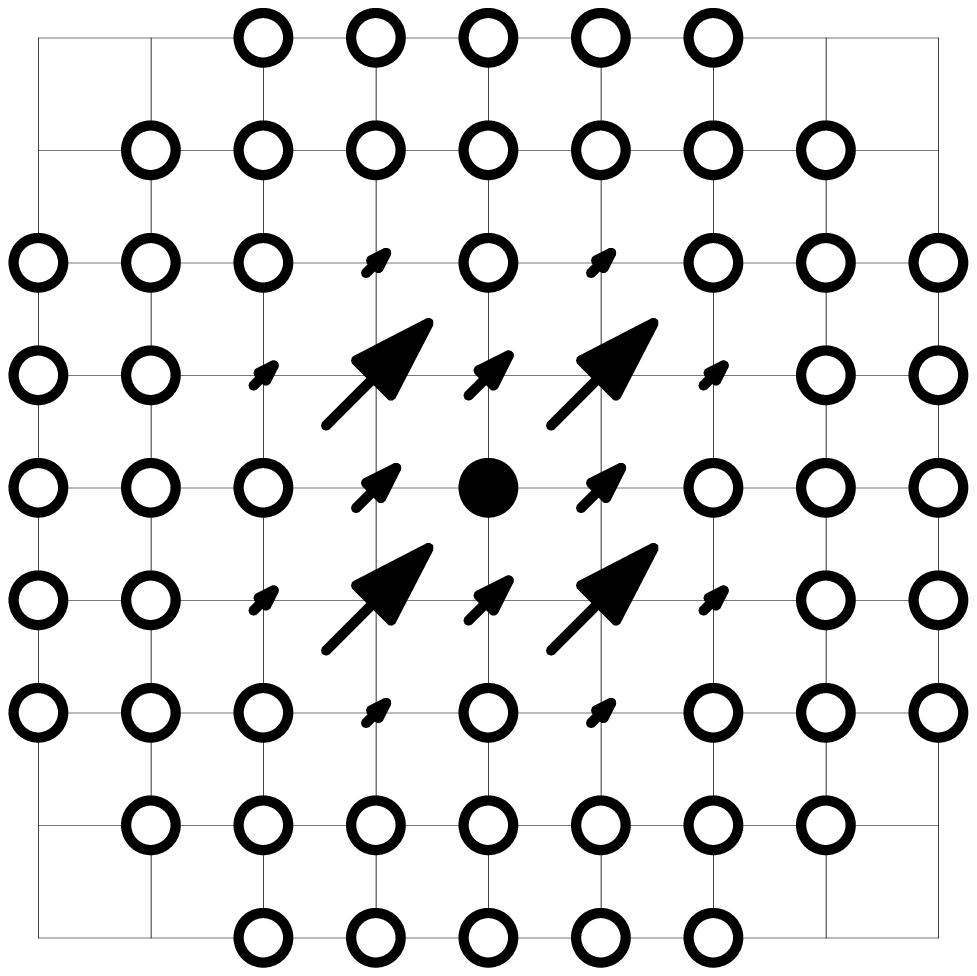}
 \label{17}}
 \subfigure[\ $ N$ =18 ]{\includegraphics[width = 27mm]{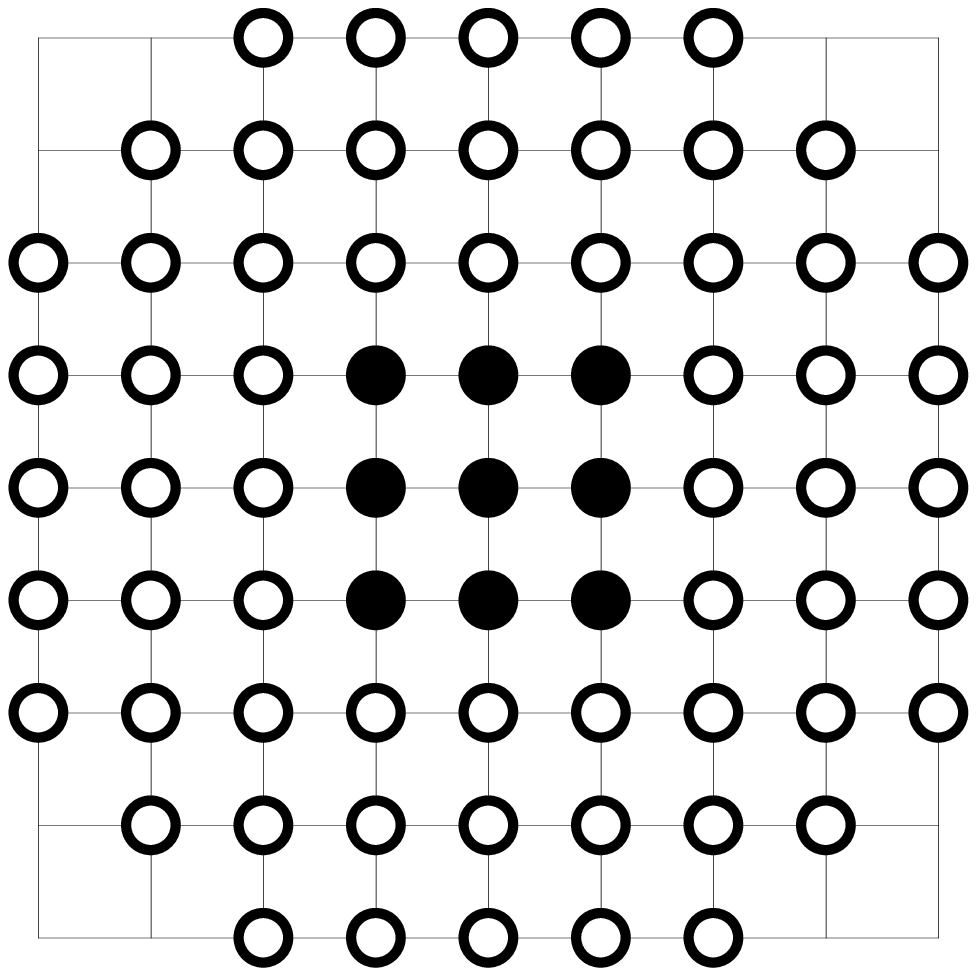}
 \label{18}}
 \subfigure[\ $ N$ =19 ]{\includegraphics[width = 27mm]{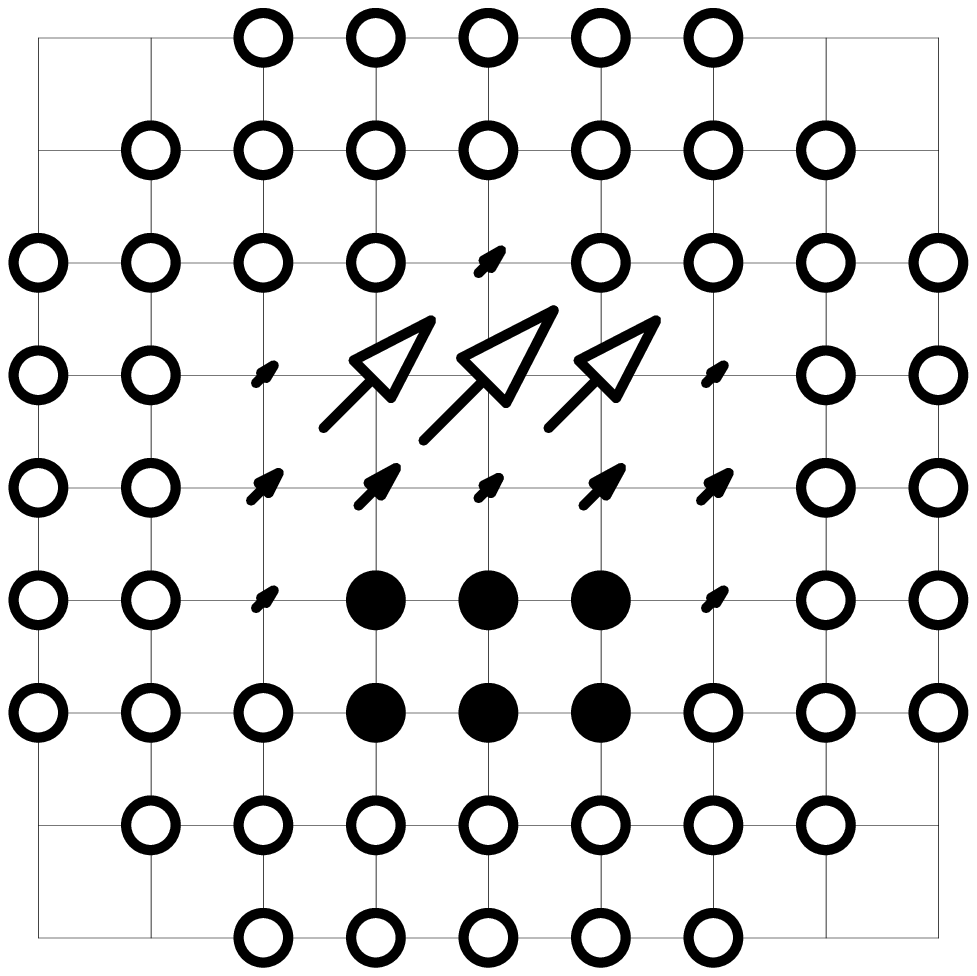}
 \label{19}}
 \caption{The distribution of the discrete variables $ \theta _i $
for a soliton for $ K/J=0.5$ at values of $ N$, close to ``magic''
value $ N=18$. Values $0\leq \theta _i <5^0$ and $175^0 <\theta _i
\leq 180^0$, corresponding to nearly maximal value of spin along the
direction up and down, presented by open and solid circles,
respectively, other notations and arrow scale here are the same as
in Fig. \ref{fig5}. For collinear structure at $ N=18$, as well as
for $ N=24$ on Fig. \ref{fig7} below, the mean spin deviation from
the nominal value $ m=\pm 1$ does not exceed $ 4\cdot 10^{-5}$. For
adjacent values of $ N$, deviation from the values $ m=\pm 1$ is not
small, namely,  value $ m\approx -0.75$ in the corners of the
soliton with $ N=17$ and $ m\approx 0.6$ in the center of a wall
segment at $ N=19$. } \label{fig6}
\end{figure}

\begin{figure}
 \subfigure[\ $ N$ =23]{\includegraphics[width = 27mm]{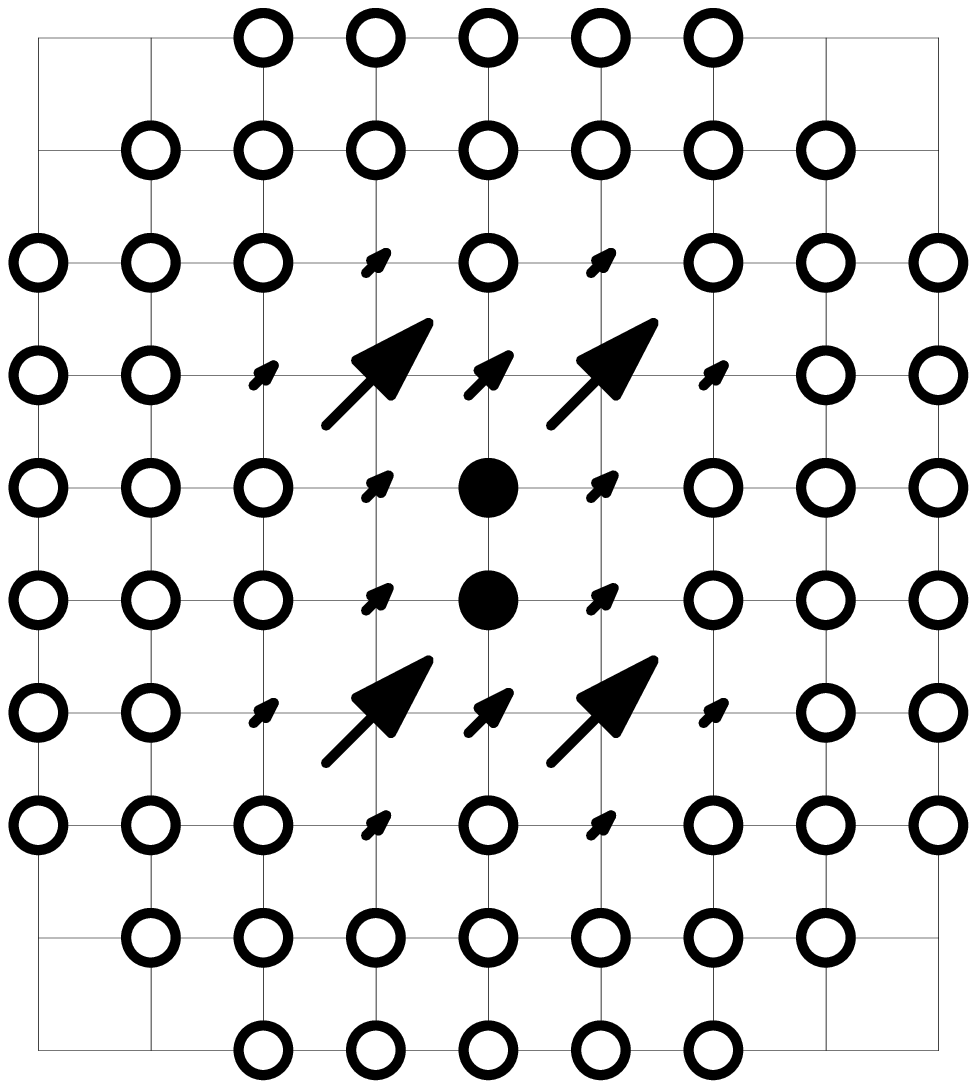}
 \label{17}}
 \subfigure[\ $ N$ =24 ]{\includegraphics[width = 27mm]{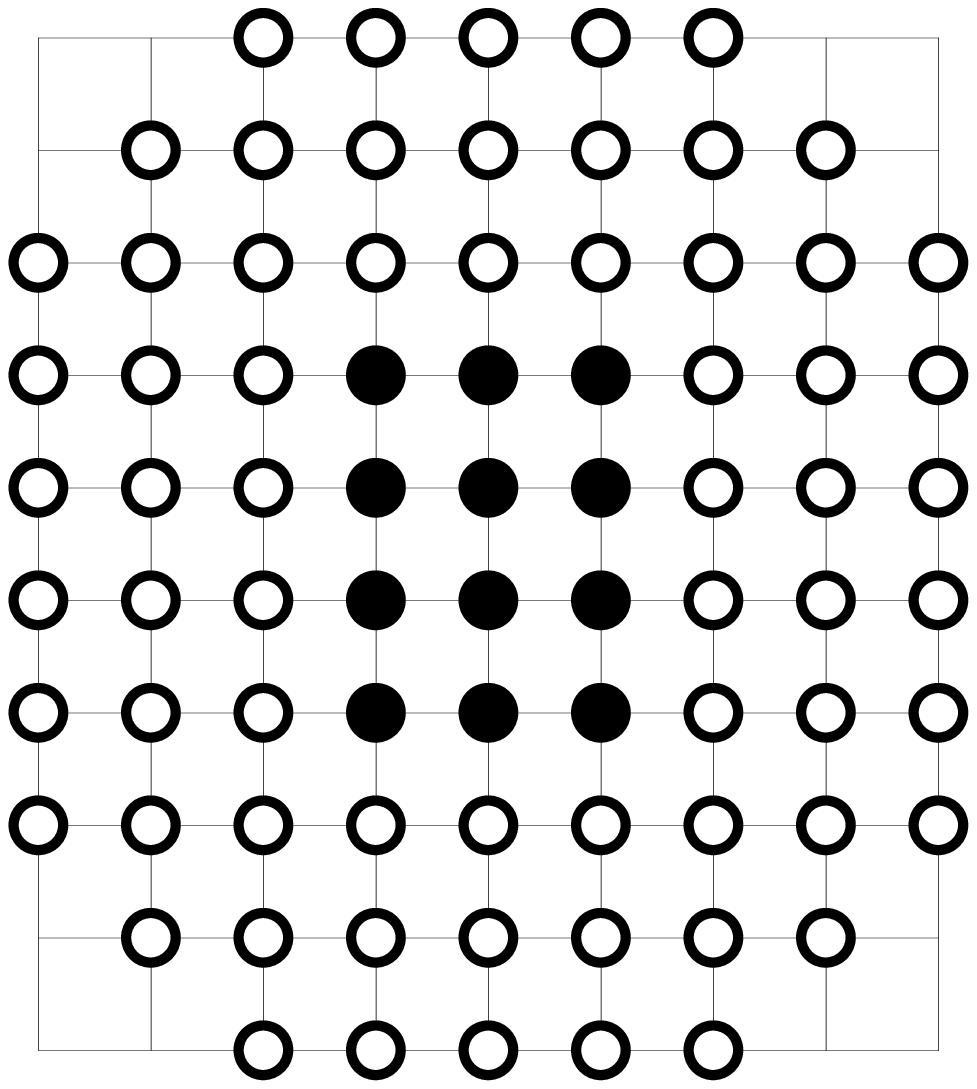}
 \label{18}}
 \subfigure[\ $ N$ =25 ]{\includegraphics[width = 27mm]{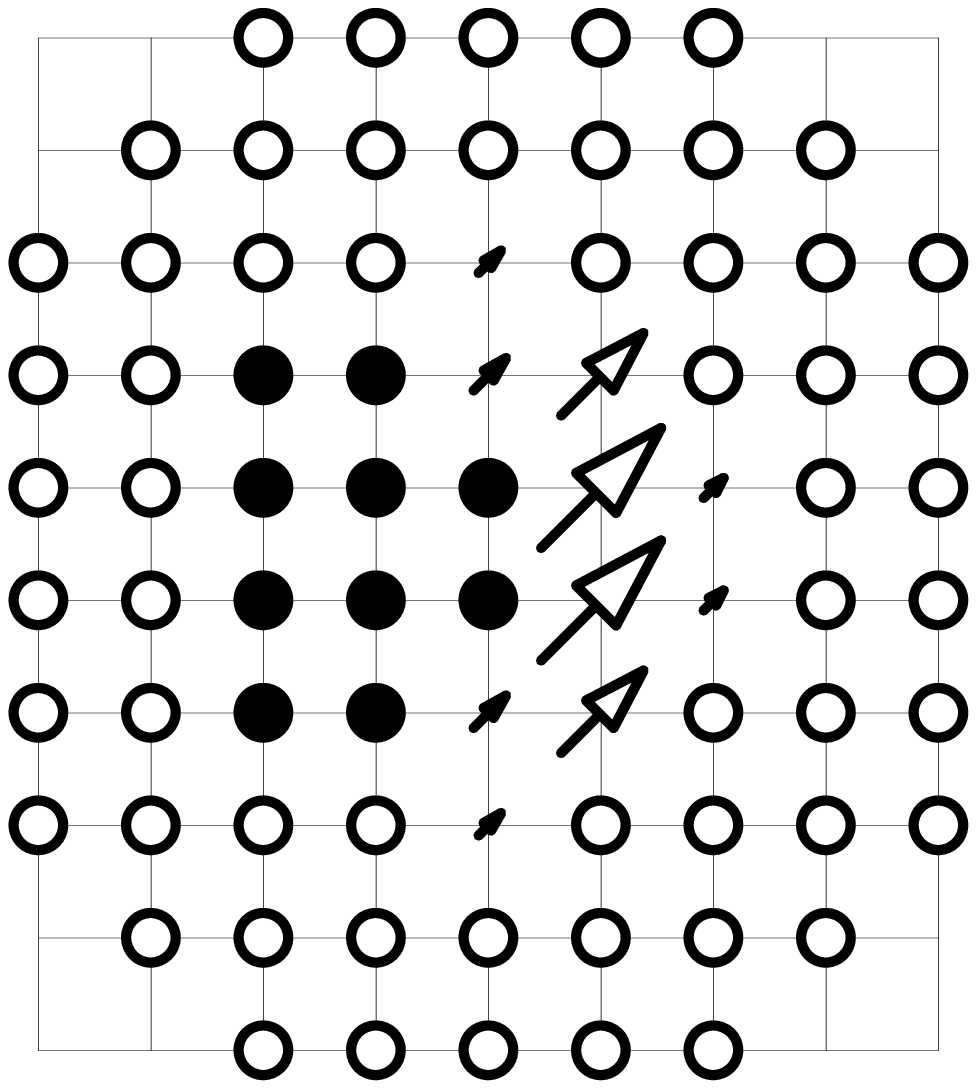}
 \label{19}}
 \caption{The distribution of the discrete variables $ \theta _i $ in a
soliton at $ N$, close to ``half-magic'' value $ N=24$. Notations,
arrow scale, and the value of $ K/J=0.5$ here are the same as in
Fig. \ref{fig6}. The value of $ m\approx -0.75$ in the corners of
the soliton with $ N=23$ and $ m\approx 0.66$ in the center of a
wall segment at $ N=25$. } \label{fig7}
\end{figure}

In the case $ r_0 \le a$ the favorable wall placed between adjacent
lines like (0,1) or (1,0) is nearly collinear. Therefore the most
favorable values of $ N$ are those for which a soliton contains a
region where all spins have $ m=-1$. The region is separated from
the rest of a magnet, where $ m=+1$, by segments of such walls. This
leads us to such spin configuration, for which this region would be
of the form of a rectangular with the size $ l_x \times l_y $, where
$l_x$ and $l_y$ are integers. Apparently, that the most preferable
would be square areas with the spin number $ l^2$ and the magnon
number $ N=2l^2$. Following,\cite{Iv+PRB06} let us call these
preferable states ``magic''. Solitons with $ N=N_{\mathrm{magic}}$
have the symmetry axis $ \mbox{C}_4 $ (hereunder we discuss the
symmetry axis perpendicular to the lattice plane), and spacious
symmetry of a soliton coincides with the lattice symmetry.

Along with magic numbers of magnons $ N_{\mathrm{magic}} =2l^2$
 ``half-magic'' states with a rectangular of spins
in which $ l_x ,\;l_y $ differ by one and $N=N_{\mathrm{half-magic}}
=2l(l+1)$ are also important. Their symmetry is lower then that for
solitons with $ N=N_{\mathrm{magic}}$ and includes only the axis $
\mbox{C}_2 $. For values like $ N=2l(l+2)$ no peculiarities were
found, most likely they are always close to magic numbers,
$l(l+2)=(l+1)^2-1$, and proximity effects to the latter are
essential.

Let us explain the character of $ \omega (N)$ in a soliton with the
value $N$, close to magic number. Configurations with
$N<N_{\mathrm{magic}} $ can be obtained by ``corner smoothing'' of a
magic configuration due to wall band inwards ``ideal square'', see
the example with $ N=17$ in Fig. 6. Since this does not require a
lot of energy, the frequency in the region $ N<N_{\mathrm{magic}} $
is small and faintly depends on $ N$, that corresponds well to our
numerical calculation. If it is necessary to increase $ N$ up to $
N>N_{\mathrm{magic}} $, then the situation is different: a domain
wall should move into a unfavorable region of the lattice. An
increase of $ N$ in the region $ N>N_{\mathrm{magic}} $ occurs only
due to this not profitable wall sector and frequency values in this
region are quite large. It is important to note that for a purely
collinear state ($ \theta =0,\;\pi )$ the change of $ \phi $ does
not change the system state, and the frequency value has no sense.
Therefore at the transition of $N$ through $ N=N_{\mathrm{magic}} $
the $ \omega (N)$ dependence experiences a jump. For this reason,
two values obtained as frequency limit at $ N=N_{\mathrm{magic}}
+\varepsilon $ and $ N=N_{\mathrm{magic}} -\varepsilon $ are
sketched in Fig.~\ref{fig4}. In real calculation the value $
\varepsilon =0.01$ was chosen.

It is worth to discuss an important problem of soliton symmetry. The
numerical analysis showed that in this case $ N \geq
N_{\mathrm{magic}} $ the wall growth occurs only from one soliton
side, see for example $ N=19$ in Fig. 6. Note the essential lowering
of the soliton symmetry for such values of $N$, for which the
soliton has no symmetry axis at all. The same regularities take
place at the transition through half-magic value, however at $
N<N_{\mathrm{half-magic}} $ the soliton symmetry is lower, it
contains an element of $ \mbox{C}_2 $ rather than $ \mbox{C}_4 $,
see Fig. 7. Then, with increasing of $N$ by values like $2 \div 3$
the soliton symmetry is restored. Further when $N$ reaches next
chosen number (half-magic after magic or vice versa) this cycle
reiterates, see Fig. 4, where the positions of magic numbers $
N=18,\;32$ and half-magic number $ N=24$ are depicted.

In fact such tendency remains at extremely small values of the ratio
$ K/J$ up to $ K/J=0.1 \div 0.3$, when $ r_0 \simeq (0.12 \div
0.23)\cdot a \ll a$. Again at $ N=1,\;2$ the soliton's size exceeds
$ a$, and the soliton can be described within the continuum
approximation. Naturally, at small $K/J$ solitons are more
localized, for example, at $ K/J=0.2$ a purely collinear states
appears at smaller $N$ like minimal magic number $ N=8$. This
quantitative difference leads to qualitatively new feature: at
values of $K < K_{\mathrm{(crit)}} \simeq 0.36J$ even the states
with $N=1$ are localized. These localized states resembles polarons,
which are one-electron states localized due to interaction with
non-linear media, see, for example.\cite{Nagaev88} These
self-localized spin states with $N=1$ can be called \emph{spin
polarons}. It is clear, the detail description of such states with
small $N$ should be based on exact quantum analysis, but we believe,
our semiclassical consideration gives at least qualitative estimate
for energies of such states. For such small $K<K_{\mathrm{(crit)}}
\ll J$ the maximal amplitude $\theta (0)$ is not small, the
asymptotical solution \eqref{eq19} is not valid more, and the energy
of spin polaron state $E^{\mathrm{sp}} \equiv E^{\mathrm{(min)}}$ is
smaller then the continual result $8(J-K)$. The value of spin
polaron energy equals to $2.57J$ at largest available value of $K =
K_{\mathrm{(crit)}} \simeq 0.36J$, and grows till $3.7J$ at $K \ll
K_{\mathrm{(crit)}}$.

\section{Conclusion remarks and result discussion. }
We considered spin dynamics in non-Heisenberg magnets with the spin
$ S=1$ and biquadratic exchange taking into account. For such
magnets, there are specific longitudinal magnetic solitons, in the
center of which the length of magnetization $ \langle S_z\rangle $
is smaller than the nominal value $ \langle S_z\rangle =1$, the
values of $ \langle S_x\rangle $ and $ \langle S_y\rangle $ are
equal to zero, but the oscillations of quadrupolar variables $
\langle S_x S_y +S_y S_x \rangle $ and $ \langle S_y^2 - S_x^2
\rangle $ are presented. The energy of such solitons is smaller then
that of standard solitons described within the Landau-Lifshitz
equation. In particular this energy is smaller then that for a
``transversal'' Belavin-Polyakov soliton, for which $ \vert
\langle{\rm {\bf S}}\rangle\vert =1$ and $ E_{BP} =4\pi J$. Note
that the Belavin-Polyakov soliton, as well as other transversal
soliton states, are also presented in the model (\ref{eq1}).

Numerical analysis of the discrete lattice model demonstrated
applicability of macroscopic approximation in vicinity of
$\mathrm{SU}(3)$ point, where $J-K \ll J$. For the rest region of
parameters, this approximation is adequate for quantitative (at
small $ N$) or semi-quantitative, in general case, description of
basic characteristics of solitons, for example, the dependence $
E(N)$. The analysis of the discrete model has also demonstrated a
series of qualitatively new effects, specifically important, when $
N$ is large enough and close to some chosen values of the magnon
number, magic $ N_{\mathrm{magic}} =2l^2$ and half-magic $
N_{\mathrm{half-magic}} =2l(l+1)$, where $ l$ is an integer. For
these chosen values a collinear state of solitons is realized, where
the magnetization in all points has maximal values $ m=\pm 1$.

This result has been obtained within the semiclassical
approximation. We were not able to construct such states within
exact quantum analysis of the model \eqref{eq1}. Let us discuss a
role of quantum effects. An important question is whether or not the
planar solution survives beyond semiclassical approximation. This
problem can be discussed by taking into account the presence of
usual ferromagnetic gapless ``transversal'' magnons within
perturbation theory. The existence of an exact semiclassical planar
solution means that the equations for transversal variables,
linearized over the planar soliton, have zero solution. In terms of
magnons, this means that processes of single magnon radiation
described by perturbation Hamiltonian $ H^{(1)}_{int}=\sum_k
[\Psi^{(1)}_{\rm {\bf p}}a^{\dag}_{\rm {\bf p}}\exp(i \omega t)
+h.c.] $ are absent (hereupon $ a_{\rm {\bf p}} ,\;a_{\rm {\bf
p}}^\dag $ are creation and annihilation operators for such magnons
with the linear momentum ${\rm {\bf p}}$ and energy $ \varepsilon
({\rm {\bf p}}) \simeq J(ap/\hbar )^2$). In principle, there is a
possibility for processes with radiation of several magnons, for
example, two-magnon process describing by $ H^{(2)}_{int}=\sum_{1,2}
(\Psi^{(2)}_{1,2} a^{\dag}_1 a^{\dag}_2\exp(i \omega t) +h.c. )$,
where $ 1\equiv {\rm {\bf p}}_1$; three-magnon process, etc.
Conservation laws of energy, momentum and total $z-$projection of
spin, which can be written as $ E(N)-E(N-2)=\varepsilon ({\rm {\bf
p}})+\varepsilon (-{\rm {\bf p}})$, allow this process even at small
$ E(N)-E(N-2) \simeq 2\hbar\omega \ll J$, as the magnon dispersion
law $\varepsilon ({\rm {\bf p}})$ has no gap. One can expect the
process of decay a soliton to magnons due to such radiations of
magnons. As a result, the soliton will be characterized by finite
lifetime. Such effects have been discussed early while going beyond
the scope of semiclassical approximation for various topological
solitons, 2D soliton with non-zero Pontryagin
index,\cite{IvSheka+PRB07} and 3D soliton characterized by non-zero
Hopf index.\cite{DzyalIv} However, for aforementioned examples these
processes are slow and the lifetime of a soliton with large enough $
N$ is long, $ \tau \sim (\hbar /J)N^5$ and $ \tau \sim (\hbar
/J)N^{5/3}$ in cases\cite{IvSheka+PRB07} and,\cite{DzyalIv}
respectively. Therefore one can expect that in our case with
complete consideration of quantum effects solitons with large value
of $N$ will be enough long-lived excitations. Detailed discussion of
this problem goes beyond of the scope of this work.

For solitons in the discrete model one can point out one more
interesting quantum effect absent at regular quantization of
continual solutions with radial symmetry. For some special numbers
of magnons $N$ the lowering of soliton symmetry $ \mbox{C}_4 $
inherent to the square lattice model (\ref{eq1}) occurs down to $
\mbox{C}_2 $ or even lower, see above Fig. 6 and 7. The presence of
solitons with symmetry lower than lattice symmetry $ \mbox{C}_4 $
means that in the classical case there are several (2 or 4)
equivalent states, which differ from each other by orientation in
the lattice. In other words the classical state of the soliton is
degenerated (two-fold or even four-fold) with regards to the soliton
orientation. In the quantum case there is a possibility for quantum
tunneling (underbarrier  transitions) between these states. For
large $N$, the transition probability is low and can be calculated
using instanton technique.\cite{IvanovFNT05,MQT} As a result one can
expect the splitting of degenerated states, with creation of doublet
or multiplet with four levels and lifting of the symmetry of the
soliton to $ \mbox{C}_4 $. We plan to return to detailed discussion
of these effects in our future work.

It is obvious that observation of effects of ``longitudinal'' spin
dynamics is possible for materials with non-small biquadratic spin
interaction. Yet Kittel demonstrated that such interaction appears
due to  interaction of spin system with lattice
deformations.\cite{Kittel} For common reason, other mechanisms like
electric multipole interactions and the Jahn-Teller effect equally
resulting in biquadratic exchange (see, e.g.,
Ref.~\onlinecite{Levy}). There are a lot of such materials widely
known, among them there are almost isotropic magnets, see review of
Nagaev.\cite{Nagaev88}

In summary it is worth to discuss a possibility of experimental
excitation of longitudinal nonlinear spin dynamics in the model
(\ref{eq1}) considered above. For a standard resonant method two
problems come up. First, frequencies of these modes are rather high;
second, magnetic field is coupled with dipole variables
(magnetization) only and does not influence directly on quadrupolar
variables. Both these problems can be solved by usage of ultrashort
intensive laser pulse, see, e.g., Refs.~\onlinecite{Kimel1,Kimel2}
and for review Ref.~\onlinecite{exper-rew}.  Usual value of a pulse
duration $ \tau $ can be as short as 100 fs, and frequencies $
\omega \ge 1/\tau $, being considerably higher than frequencies of
regular spin oscillations can be effectively excited.

The possible role of different variables, dipolar and quadrupolar,
can be demonstrated by a simple example. Consider a thick plane --
parallel plate of a ferromagnet saturated along its normal ($z$
axis). Let the light pulse propagates along the $ z$ axis, with the
electric field parallel to the plate surface. The light interaction
with dipolar degrees of freedom $ m_i =\langle S_i \rangle$ can be
described as following. Due to inverse Faraday effect, a circularly
polarized light is equivalent to pulse magnetic field parallel to $
z$ axis.\cite{Kimel1,Kimel2} Linearly polarized light produces
two-fold anisotropy in the sample's plane.\cite{PulseGarnets} Both
scenario are ineffective for a sample, saturated along $z-$axis, and
the excitation of usual transversal spin oscillations (magnons) is
absent in this geometry.

In contrast, the quadrupolar variables like $ \langle S_i S_j +S_j
S_i\rangle $ with $ i,j=x,y$ are coupled directly with linearly
polarized light. The influence of such light pulse is equivalent to
the \textit{direct action} of some pulse of effective magnetic field
$ {\rm {\bf H}}^Q(t)$, see Eq. (\ref{eq14}), on the variable
$\boldsymbol{\sigma}$ of the form $ H^{(int)}=-{\rm {\bf
H}}^Q\boldsymbol{\sigma}$. Being directed perpendicularly to the
``ground state magnetization'' $ \boldsymbol{\sigma} =
\mathbf{e}_z$, this pulse field effectively excite the oscillations
of the $x$ and $y$ components of $\boldsymbol{\sigma}$, that is, the
longitudinal spin oscillations considered in this article. The
excitation of non-dipolar spin degrees of freedom by use of
ultrafast optical pumping was recently observed for magnetic Mott
insulator R$_2 $CuO$ _4$.\cite{Pisarev+}

We are thankful to V.~G. Bar'yakhtar, A.~K. Kolezhuk and D.~D. Sheka
for stimulated discussions. The work is partially supported by the
grant INTAS-05-1000008-8112 and by the joint grant from Ministry of
Education and Science of Ukraine and Ukrainian State Foundation of
Fundamental Research F25.2/081.

\end{document}